\documentclass[prl,twocolumn,aps,floatfix,longbibliography,superscriptaddress]{revtex4-2}
\usepackage[utf8]{inputenc}
\usepackage{natbib}
\usepackage[normalem]{ulem}
\usepackage{graphicx}
\usepackage{xcolor}

\usepackage[tbtags]{amsmath}
\usepackage{hyperref}
\hypersetup{colorlinks,allcolors=black}
\usepackage{amssymb}
\usepackage{gensymb}
\usepackage{float}
\usepackage{amsmath}
\usepackage{tabularx,graphicx}
\usepackage{epstopdf}
\usepackage{latexsym}
\usepackage{color, colortbl}
\usepackage{psfrag}
\usepackage{bbm}
\usepackage{bm}
\usepackage{titlesec}
\usepackage{dsfont}
\usepackage{feynmp}
\usepackage{slashed}
\usepackage{multirow}
\usepackage{soul} 

\newcommand{\br}{\bm{r}}
\newcommand{\bk}{\bm{k}}
\newcommand{\bq}{\bm{q}}

\newcommand{\be}{\begin{equation}}
\newcommand{\ee}{\end{equation}}
\newcommand{\beq}{\begin{eqnarray}}
\newcommand{\eeq}{\end{eqnarray}}
\newcommand{\ba}{\[\begin{aligned}}
\newcommand{\ea}{\end{aligned}\]}

\renewcommand{\phi}{\varphi}
\renewcommand{\epsilon}{\varepsilon}
\renewcommand{\dag}{\dagger}
\newcommand{\SC}{{\mathrm{SC}}}
\newcommand{\nt}{\notag\\}

\definecolor{orange}{rgb}{1,0.5,0}

\hypersetup{colorlinks=true,linkcolor=blue,citecolor=blue,urlcolor=blue}

\begin{document}
\title{Andreev reflection spectroscopy in strongly paired superconductors}

\author{Cyprian Lewandowski}
\affiliation{Department of Physics, California Institute of Technology, Pasadena CA 91125, USA}
\affiliation{Institute for Quantum Information and Matter, California Institute of Technology, Pasadena CA 91125, USA}
\email[]{cyprian@caltech.edu}

\author{\'Etienne Lantagne-Hurtubise}
\affiliation{Department of Physics, California Institute of Technology, Pasadena CA 91125, USA}
\affiliation{Institute for Quantum Information and Matter, California Institute of Technology, Pasadena CA 91125, USA}

\author{Alex Thomson}
\affiliation{Department of Physics, California Institute of Technology, Pasadena CA 91125, USA}
\affiliation{Institute for Quantum Information and Matter, California Institute of Technology, Pasadena CA 91125, USA}
\affiliation{Walter Burke Institute for Theoretical Physics, California Institute of Technology, Pasadena, California 91125, USA}
\affiliation{Department of Physics, University of California, Davis, California 95616, USA}

\author{Stevan Nadj-Perge}
\affiliation{T. J. Watson Laboratory of Applied Physics, California Institute of
  Technology, 1200 East California Boulevard, Pasadena, California 91125, USA}
\affiliation{Institute for Quantum Information and Matter, California Institute of Technology, Pasadena CA 91125, USA}

\author{Jason Alicea}
\affiliation{Department of Physics, California Institute of Technology, Pasadena CA 91125, USA}
\affiliation{Institute for Quantum Information and Matter, California Institute of Technology, Pasadena CA 91125, USA}

\begin{abstract}
Motivated by recent experiments on low-carrier-density superconductors, including twisted multilayer graphene, we study signatures of the BCS to BEC evolution in Andreev reflection spectroscopy. We establish that in a standard quantum point contact geometry, Andreev reflection in a BEC superconductor is unable to mediate a zero-bias  conductance beyond  $e^2/h$ per lead channel.  
This bound is shown to result from a duality that links the sub-gap conductance of BCS and BEC superconductors. We then demonstrate that sharp signatures of BEC superconductivity, including perfect Andreev reflection, can be recovered
by tunneling through a suitably designed potential well. We propose various tunneling spectroscopy setups to experimentally probe this recovery.
\end{abstract}

\maketitle

\begin{figure}[t]
	\includegraphics[width=\columnwidth]{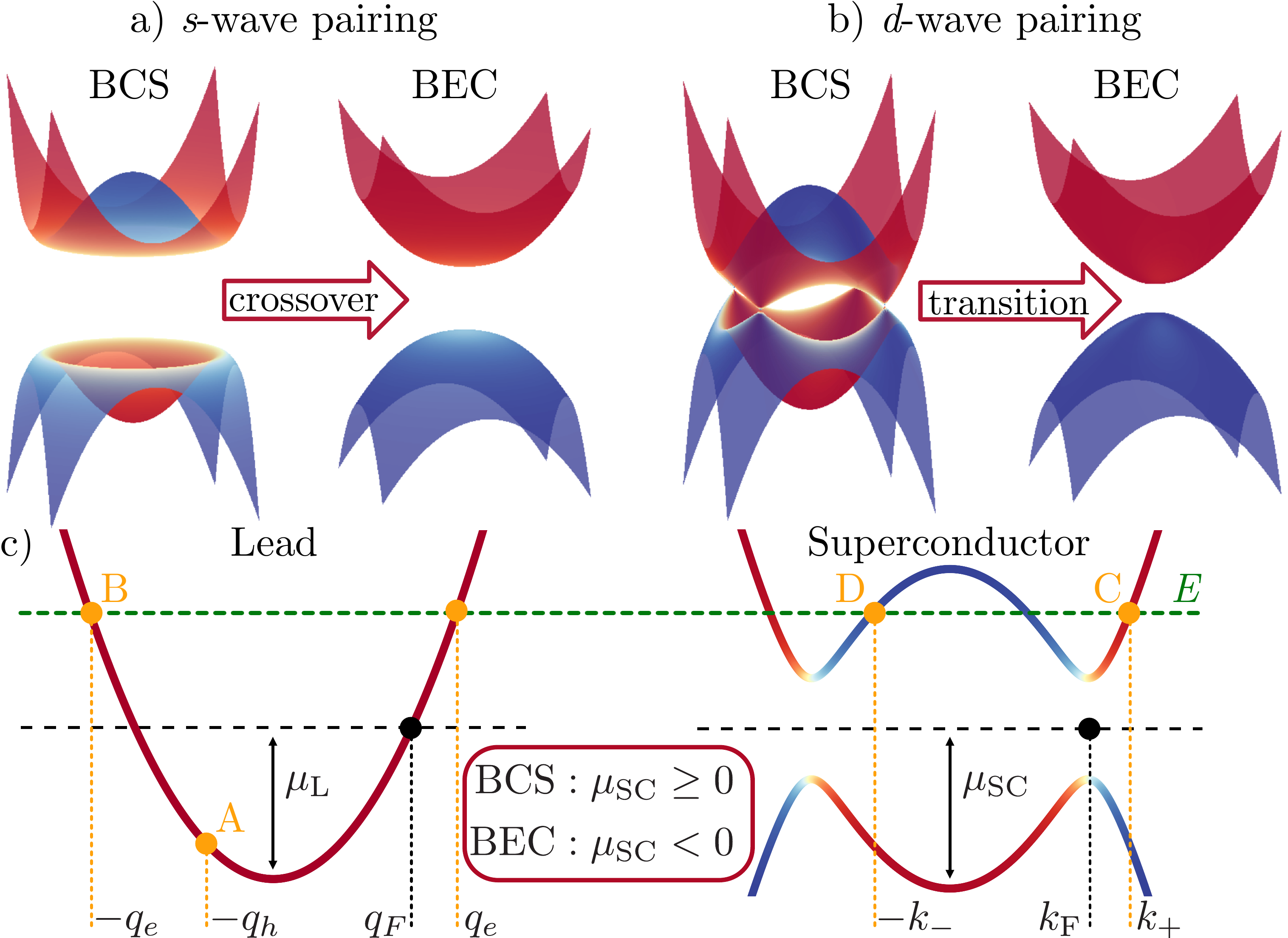}
	\caption{(a,b) Quasiparticle spectra across the BCS-to-BEC superconductor evolution, which is (a) a crossover for $s$-wave pairing but (b) a Lifshitz transition for nodal (e.g., $d$-wave) pairing. (c) Scattering processes at the normal-superconductor interface included in the BTK formalism: Andreev (A) and normal (B) reflection, and direct (C) and band-crossing (D) transmission as a quasiparticle. The green dashed line denotes the incident electron energy.}
	\label{fig1}
\end{figure}

The evolution from Bardeen-Cooper-Schrieffer (BCS) to Bose-Einstein condensate (BEC) superconductivity has been a recurring theme in the study of physical systems ranging from cold atomic gases to
strongly correlated materials and neutron stars~\cite{Chen2005, DeMelo2008, Zwerger2011, Randeria2014, Strinati2018}. 
In the BCS regime, superconductivity arises from weakly bound Cooper pairs with a characteristic size $\xi_\mathrm{pair}$ that far exceeds the mean inter-particle spacing $d$; in the BEC regime, by contrast, fermions form tightly bound Cooper pairs with size $\xi_\mathrm{pair} \ll d$. BCS- and BEC-type superconductors can be separated either by
a crossover (e.g., for $s$-wave pairing~\cite{Eagles1969, Leggett1980, Nozieres1985, Randeria1989, deMelo1993, Pistolesi1996};  Fig~\ref{fig1}a), a Lifshitz-type phase transition (e.g., for nodal pairing~\cite{Randeria1990, Duncan2000, Borkowski2001}; Fig.~\ref{fig1}b), or a topological phase transition~\cite{Read2000}.

Ultracold atoms, where the pairing strength between fermions can be continuously tuned 
with the aid of a Feshbach resonance~\cite{Regal2004, Zwierlein2004, Chin2004, Bourdel2004, Partridge2005, Zwierlein2005}, provide a controlled experimental platform for exploring the evolution from weak to strong pairing. While most solid-state superconductors reside firmly in the BCS regime, certain strongly correlated materials such as cuprates have been rationalized in terms of the BCS-BEC paradigm~\cite{Harrison2022}. Further, a new generation of experiments probing low-carrier-density materials including iron-based compounds~\cite{Kasahara2014, Rinott2017, Hashimoto2020,Mizukami2021}, Li$_x$ZrNCl~\cite{Nakagawa2021, Shi2021} and moir\'e graphene systems
\cite{Park2021, Hao2021, Kim2021, tian2021, Park2021b, Zhang2021} has revealed signatures consistent with proximity to a BEC state---opening a new experimental frontier for unconventional superconductivity. Developing probes that can unambiguously identify BEC superconductors and distinguish possible competing phases therefore poses a pressing problem. 

In solid-state contexts, BEC superconductivity can manifest in various ways.  Saturation of the Ginzburg-Landau coherence length $\xi_{\rm GL}$ at the interparticle spacing~\cite{Nakagawa2021,Park2021} and a critical temperature approaching theoretical bounds~\cite{Botelho2006, Hazra2019} both constitute indirect evidence for proximity to the strong pairing regime. 
 Energy-momentum resolved probes~\cite{Rinott2017} can also reveal the evolution of the quasiparticle dispersion from BCS-like to BEC-like (Fig.~\ref{fig1}a,b). Superconductors with a nodal order parameter exhibit a Lifshitz transition from a gapless BCS to a gapped BEC phase, which leads to a predicted divergence in electronic compressibility at the transition~\cite{Duncan2000} and a gap opening that can be observed, e.g., via scanning tunneling microscopy (STM)~\cite{Randeria1990, Borkowski2001, Kim2021}. In STM it is, however, difficult to distinguish BEC superconductors from competing insulators since, unlike their BCS counterparts, they feature less prominent and particle-hole asymmetric coherence peaks~\cite{Loh2016, Kim2021}.
 
 \begin{figure*}[t]
	\includegraphics[width=\textwidth]{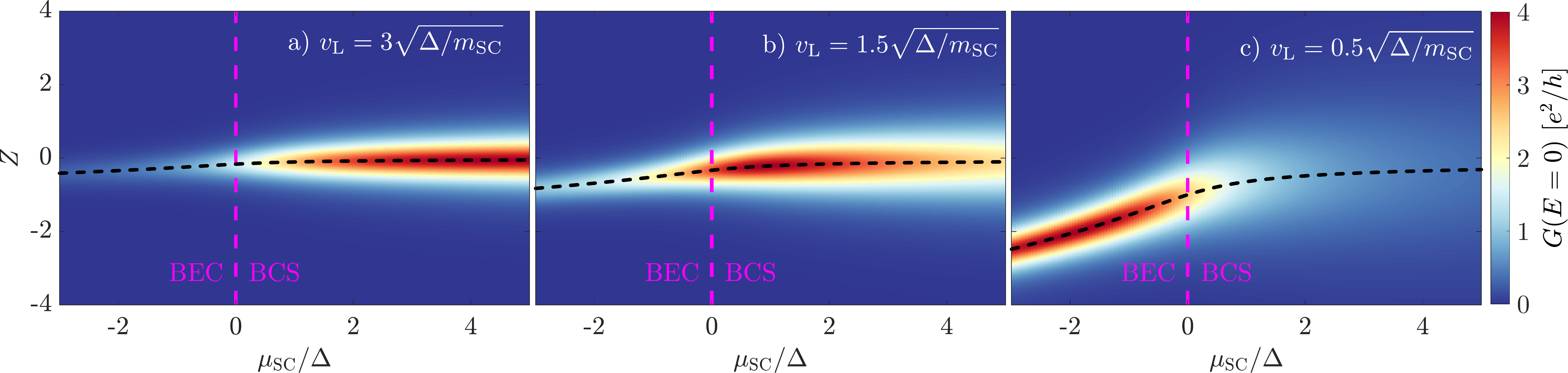}
	\caption{Dependence of the BTK zero-bias tunneling conductance on the barrier parameter $Z$ along the BCS-to-BEC crossover for a one-dimensional $s$-wave superconductor with $v_L \sqrt{m_{\rm SC}/\Delta}$ set to (a) 3, (b) 1.5, and (c) 0.5.
	Black dashed lines trace $Z = -v_{\rm R}/2 v_{\rm L}$, one of the necessary conditions for perfect Andreev reflection.
	Incoming electrons tunnel through a potential barrier at $Z>0$ but a potential well at $Z<0$.  While the conductance in the BCS regime is approximately $Z\rightarrow -Z$ symmetric, pronounced asymmetry emerges in the BEC regime, with enhanced Andreev reflection possible at negative $Z$ [see panel (c)].
	}
	\label{fig2}
\end{figure*}

Here we investigate Andreev reflection spectroscopy across the BCS-BEC evolution and predict striking new manifestations of BEC superconductivity. First, in a standard quantum point contact (QPC) geometry~\cite{Daghero2010}
we find that Andreev reflection is suppressed upon passing from the BCS to BEC regime,
consistent with the analysis of BEC-BEC Josephson junctions in Ref.~\citenum{Setiawan2021}. 
We show that this suppression is a consequence of a duality that interchanges
the wavefunctions of BEC and BCS superconductors. Second, we establish that Andreev reflection can be controllably revived in the BEC regime by tunneling between the lead and superconductor through an effective potential \emph{well}---as opposed to the barrier employed in conventional treatments~\cite{BTK1982}.  This feature sharply distinguishes BEC from BCS superconductors and can be probed in experimental setups that we propose.  

{\bf \emph{Setup.}}~We first employ the Blonder-Tinkham-Klapwijk (BTK) framework, which describes scattering between a normal lead and a superconductor~\cite{BTK1982}. Consider a 1D setup (Fig.~\ref{fig1}c) with an interface at position $x=0$ separating a spinful normal lead (at $x<0$) with dispersion $\xi_q = \frac{q^2}{2m_{\rm L}} - \mu_{\rm L}$ from a singlet superconductor (at $x>0$) with dispersion $\xi_k = \frac{k^2}{2m_{\rm SC}} - \mu_{\rm SC}$ and $s$-wave pairing potential $\Delta$. We take $\mu_{\rm L}>0$ throughout but allow $\mu_{\rm SC}$ to take either sign to capture the BCS-BEC crossover that occurs as $\mu_{\rm SC}$ crosses zero. A delta-function tunnel barrier, $\lambda \delta(x)$, interpolates between the tunneling limit with large $\lambda > 0$ and the QPC limit $\lambda=0$.

Figure~\ref{fig1}c illustrates the processes available to an incident electron at the interface: Andreev (A) or normal (B) reflection to a hole or electron, respectively, and transmission to an electron- (C) or hole-like (D) quasiparticle. The probabilities for these processes are obtained by matching the wavefunctions and their first derivatives at the interface and normalizing by the appropriate group velocities~\cite{SM}. In the standard BTK formalism applied to weakly paired BCS superconductors, the limit $\Delta, E \ll \mu_{\rm L} , \mu_{\rm SC}$ combined with the equal-mass assumption $m_{\rm L} = m_{\rm SC}$ drastically simplifies the problem, yielding the Andreev approximation in which $q_h = q_e = k_\pm = q_F$ with $q_F = \sqrt{2 m_{\rm L} \mu_{\rm L}}$ the lead's Fermi momentum~\cite{andreev1964}. To describe the BCS-BEC evolution, we instead treat the problem in full generality  (see Ref.~\onlinecite{Setiawan2021} for an analogous treatment of Josephson junctions).
The tunneling conductance $G(E) = dI/dV$ at bias energy $E = eV$
is given in the Landauer-B\"uttiker formalism by 
\begin{equation}
\label{eq:buttiker}
    G(E) = \frac{2e^2}{h}\left[1+ A(E) - B(E)\right],
\end{equation}
where $A(E)$ and $B(E)$ denote the Andreev and normal reflection probabilities. Maximal conductance of $4 e^2/h$ corresponds to perfect Andreev reflection $A(E)=1$.

{\bf \emph{Zero-bias conductance.}}~For gapped superconductors, the absence of transmission processes (C and D) considerably simplifies the analysis of the sub-gap conductance: employing the normalization condition $A(E) +B(E) = 1$ returns $G(E) = \frac{4e^2}{h}A(E)$.  
As shown in the Supplement~\cite{SM}, the zero-bias Andreev reflection probability, $A_0 \equiv A(E=0)$, then reduces to 
\begin{equation}
    A_0 = \frac{ 4 v_{\rm L}^2 v_{\rm I}^2 }{\left[\left(2 v_{\rm L} Z + v_{\rm R}\right)^2 + v_{\rm I}^2 + v_{\rm L}^2\right]^2},
    \label{eq:BTK_finiteZ}
\end{equation}
where $Z = \lambda/v_{\rm L}$ quantifies the barrier transparency, $v_{\rm L} = q_F/m_{\rm L}$ is the Fermi velocity of the lead, and $v_{\rm R, I} = \kappa_{\rm R,I}/m_{\rm SC}$ are two characteristic velocities of the superconductor. The momenta $\kappa_{\rm R, I}$ are defined through
\begin{equation}
\kappa e^{i \phi} = \kappa_{\rm R} + i \kappa_{\rm I} = \sqrt{2 m_{\rm SC} (- \mu_{\rm SC} + i \Delta) },
\label{eq:kappa_definition}
\end{equation}
and the length scales $\kappa_{\rm R}^{-1}$ and $\kappa_{\rm I}^{-1}$ respectively
control the evanescent and oscillatory behavior of the wavefunction in the superconductor at $x>0$, $\psi_\SC\sim e^{-(\kappa_{\rm R}+i\kappa_{\rm I})x}$. 

Sending $\mu_{\rm SC} \rightarrow -\mu_{\rm SC}$, which connects the BCS and BEC regimes, yields $\varphi \rightarrow \pi/2 - \varphi$ and hence swaps $\kappa_{\rm R} \leftrightarrow \kappa_{\rm I}$; cf.~Eq.~\eqref{eq:kappa_definition}.  This remarkable property reveals a duality between the BCS and BEC superconductor wavefunctions.  
As an instructive limiting case, deep in either the BEC or BCS regimes where $|\mu_\mathrm{SC}|/\Delta\gg1$,  $\kappa_{\rm R,I}$ can be expanded as
\begin{align}\label{eqn:kappa_duality}
    \kappa_{\rm R}+i\kappa_{\rm I}
    &\approx 
    \sqrt{\mathrm{sgn}(\mu_\mathrm{SC})}\left(\frac{\Delta}{v_F}+ i \mathrm{sgn}(\mu_\mathrm{SC})k_F \right).
\end{align}
Here the definition of the Fermi momentum $k_F=\sqrt{2m_\mathrm{SC}|\mu_\SC|}$ and velocity $v_F=k_F/m_\SC$ is extended into the BEC regime.
In the BCS limit Eq.~\eqref{eqn:kappa_duality} aligns with conventional understanding: The wavefunction's oscillatory part is set by $k_F$, whereas evanescent decay is controlled by $\Delta/v_F$---i.e., the inverse of the BCS pair coherence length $\xi_{\rm pair}$. The situation flips in the BEC regime, where
$k_F$ controls the decay length and $\Delta/v_F$ dictates the oscillatory behavior. 

This duality has immediate implications for the conductance. In the QPC limit, $Z=0$, the Andreev reflection probability in Eq.~\eqref{eq:BTK_finiteZ} is maximized when the lead Fermi velocity satisfies $v_{\rm L} =\sqrt{v_{\rm R}^2 + v_{\rm I}^2}$, yielding the upper bound $A_0 \leq v_{\rm I}^2/(v_{\rm I}^2+v_{\rm R}^2)$ valid for any $\mu_{\rm SC}$. Near-perfect Andreev reflection thus requires $v_{\rm R} \ll v_{\rm I}$, implying many oscillation periods over the decaying envelope of the superconductor wavefunction. This requirement exactly translates to the BCS limit, $\mu_\SC\gg\Delta$, as originally pointed out by Andreev~\cite{andreev1964}.
By contrast, the deep BEC regime is characterized by the converse inequality, $v_{\rm R} \gg v_{\rm I}$, an the upper bound above accordingly reduces to $A_0 \lesssim v_{\rm I}^2/v_{\rm R}^2=\Delta^2/\mu_\SC^2$. Physically, the fast decay of the superconducting wavefunction relative to its oscillation period suppresses Andreev reflection from the plane wave electrons of the lead. At the ``self-dual" point $\mu_{\rm SC} = 0$, the superconductor exhibits a single length scale, with $v_{\rm R} = v_{\rm I} = \sqrt{\Delta/m_{\rm SC}}$, yielding an Andreev reflection probability upper-bounded by $1/2$.

{\bf \emph{Andreev revival.}}~The preceding bound on Andreev reflection in the BEC regime ($\mu_{\rm SC} < 0$) implies that the zero-bias conductance in the QPC limit cannot exceed the value $2e^2/h$ characteristic of a perfect metallic contact. According to Eq.~\eqref{eq:BTK_finiteZ}, moving away from the QPC limit by adding a tunnel barrier modeled by $Z>0$ only further suppresses the Andreev reflection probability $A_0$. 
Interestingly, however, Andreev reflection can be enhanced by exploiting a pronounced asymmetry in the sign of $Z$ that emerges in the BEC regime.

Writing $v_{\rm I} = \alpha v_{\rm L}$ and $v_{\rm R} = \beta v_{\rm L}$, one can re-express Eq.~\eqref{eq:BTK_finiteZ} as $A_0 = 1/(1 + 2\tilde{Z}^2)^2$ in terms of a renormalized transparency parameter,
\begin{align}
   \tilde{Z}^2 &= \frac{(Z + \beta/2)^2}{\alpha} + \frac{(\alpha-1)^2}{4 \alpha},
   \label{eq:tilde_Z}
\end{align}
that incorporates both the intrinsic barrier transparency $Z$ and velocity mismatch effects~\cite{SM}.
The linear-in-$Z$ contribution in Eq.~\eqref{eq:tilde_Z} is proportional to the dimensionless parameter $\beta/\alpha = v_{\rm R}/v_{\rm I}$ controlling the BEC to BCS evolution. 
In the BCS limit, the distinction between $Z$ positive and negative---i.e., potential barriers and wells---is correspondingly negligible since $v_{\rm I}  \gg v_{\rm R}$. In the BEC regime, by contrast,
$v_{\rm R} > v_{\rm I}$
implies sensitive dependence on the sign of $Z$. 
Indeed, potential wells characterized by $Z<0$ can \emph{promote} Andreev reflection---which becomes perfect at zero bias even deep within the BEC regime when $\tilde{Z} = 0$, i.e., when $Z = - v_{\rm R}/(2v_{\rm L})$ and $v_{\rm L} = v_{\rm I}$. These trends are illustrated in Fig.~\ref{fig2}, which plots the zero-bias conductance versus $Z$ across the BCS-BEC evolution tuned by $\mu_{\rm SC}/\Delta$, for different lead velocities $v_{\rm L}$ in panels (a)-(c).
Note that the standard BTK result for a BCS superconductor is recovered for large $v_{\rm L}$ [Fig.~\ref{fig2}(a)].

\begin{figure}[t]
	\includegraphics[width=0.8\linewidth]{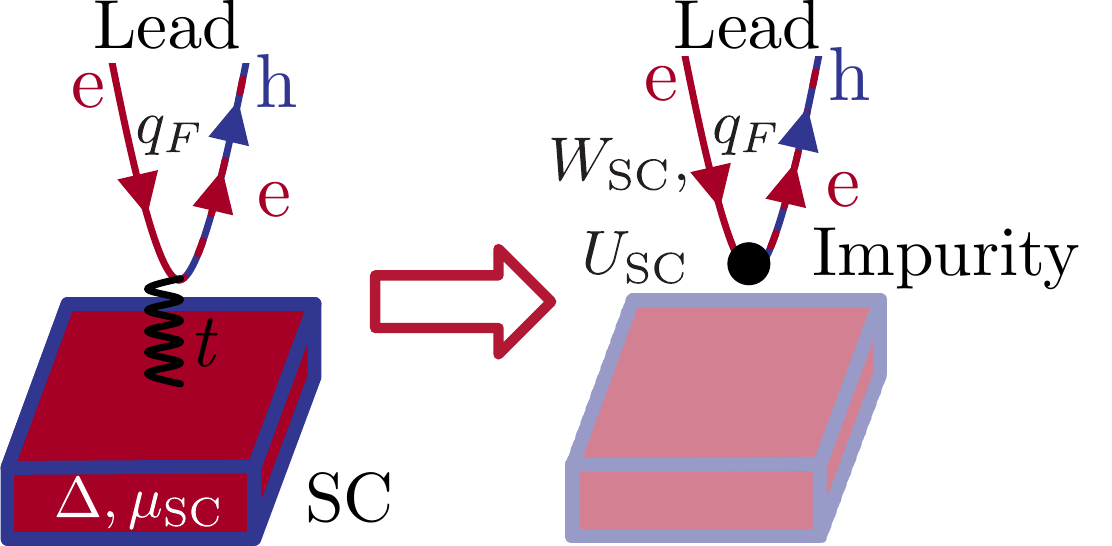}
	\caption{Sketch of the effective lead formalism.  A lead tunnels electrons onto a gapped superconductor (left). Integrating out the gapped degrees of freedom generates an effective lead-only ``impurity'' problem (right) featuring a shifted
	local potential $U_{\rm SC}$ and a local pairing term $W_{\rm SC}$ mediated by the superconductor.
	}
	\label{fig3}
\end{figure}

\begin{figure*}[t]
	\includegraphics[width=\textwidth]{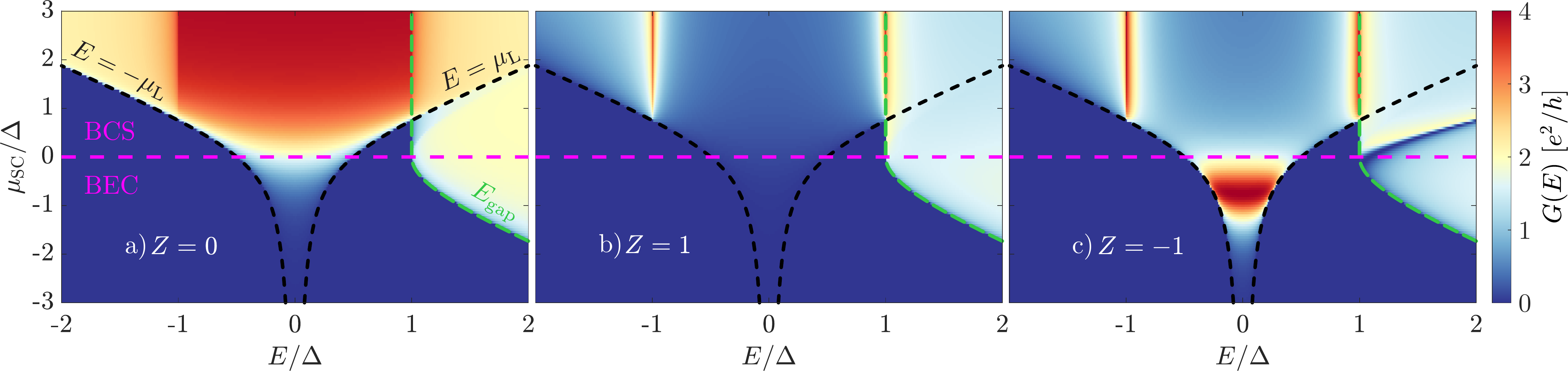}
	\caption{Finite-bias tunneling spectroscopy across the BCS-BEC evolution with (a) $Z=0$, (b) $Z=1$ and (c) $Z=-1$, assuming an optimal lead with $v_{\rm L} = v_{\rm I}$ for each $\mu_{\rm SC}$ value.
	Black dashed lines at energies $E = \pm \mu_{\mathrm L}$ show the corresponding optimal lead chemical potential $\mu_{\rm L}$, and green dashed lines denotes the quasiparticle gap $E_{\rm gap}$. (a) The quantum point contact limit $Z=0$ exhibits a plateau-like enhancement of sub-gap conductance due to Andreev reflection in the BCS regime, but not in the BEC regime. (b,c) Tunneling spectra for $Z = \pm 1$ are similar in the BCS regime, showing a gap surrounded by coherence peaks at $E=\pm \Delta$. In the BEC regime, sub-gap conductance enhancement occurs \emph{only} for tunneling through potential wells, as in (c).}
	\label{fig4}
\end{figure*}

{\bf \emph{Effective lead formalism.}}~For complementary insight,
we now examine
normal metal-superconductor tunneling via an effective lead formalism (ELF).
Consider a 1D lead at positions $x\leq 0$ with linearized kinetic energy and Fermi velocity $v_\mathrm{L}=q_F/m_\mathrm{L}$.  At its endpoint, the lead  
exhibits a local potential $U_0\delta(x)$ and couples to a gapped $d$-dimensional superconductor via electron hopping of strength $t$; see Fig.~\ref{fig3}. 
Focusing on sub-gap energies, the superconductor's degrees of freedom can be safely integrated out---yielding an effective lead-only Hamiltonian with  additional (marginal) ``impurity'' perturbations \cite{SM}.
Specifically, the superconductor both shifts the local potential $U_0$ by 
\begin{equation}
    U_{\rm SC}
    = t^2 \int \frac{d^d \bk}{(2 \pi)^d} \frac{\xi_{\bk}}{\xi_{\bk}^2 + \Delta^2 }
\end{equation}
and generates a local singlet pairing term with amplitude   
\begin{align}
    W_{\rm SC}
     &= t^2 \int \frac{d^d \bk}{(2 \pi)^d} \frac{\Delta}{\xi_{\bk}^2 + \Delta^2 } .
\label{eq:ELF_parameters}
\end{align}
Explicit evaluation of the integrals reveals that, for $d = 1$, $U_{\rm SC}$ and $W_{\rm SC}$ exactly swap under $\mu_{\rm SC} \rightarrow -\mu_{\rm SC}$, manifesting the duality uncovered above within the BTK formalism (for details and extensions see \cite{SM}).

Extracting the wavefunctions from the effective lead-only Hamiltonian yields a zero-bias Andreev reflection probability  \cite{SM}
\begin{equation}
\label{eq:ELF_andreev}
      A_0^{\mathrm{ELF}} = \frac{4 v_{\rm L}^2 (W_{\rm SC}/2)^2}{\left[ (U_\mathrm{eff}/2)^2 + (W_\mathrm{SC}/2)^2 +v_{\rm L}^2\right]^2},
\end{equation}
with $U_{\rm eff} = U_0 + U_{\rm SC}$. Eq.~\eqref{eq:ELF_andreev}  reproduces Eq.~\eqref{eq:BTK_finiteZ} upon identifying the ELF/BTK correspondence $U_0/2 \leftrightarrow 2 v_{\rm L} Z$, $U_{\rm SC}/2 \leftrightarrow v_{\rm R}$, and $W_{\rm SC}/2 \leftrightarrow v_{\rm I}$.
Thus negative values of $U_0$ enable the cancellation of the `large' potential $U_{\rm SC}$ generated by BEC superconductors---in turn allowing Andreev processes mediated by $W_{\rm SC}$ to dominate.

{\bf \emph{Extensions.}}~Figure~\ref{fig4} presents the finite-bias conductance $G(E)$ for  an $s$-wave superconductor.  Data were obtained numerically from the full BTK analysis, including quasiparticle transmission channels, for (a) $Z = 0$, (b) $Z = 1$, and (c) $Z = -1$.  To clearly illustrate the revival of Andreev reflection, the lead Fermi velocity is tuned such that $v_{\rm L} = v_{\rm I}$ across the BCS-BEC evolution; that is, for each $\mu_{\rm SC}$ on the vertical axis we fix the lead chemical potential to its optimal value (see dashed black curves). In the QPC limit, $Z=0$, Fig.~\ref{fig4}(a) shows the familiar sub-gap conductance plateau at $G(E) = 4e^2/h$ in the BCS regime~\cite{BTK1982}. Upon entering the BEC regime, the entire plateau is suppressed in accordance with the bound on Andreev reflection at zero bias derived above. In the tunneling limit ($Z>0$),
the BCS regime exhibits sharp coherence peaks at the gap edge that similarly diminish upon entering the BEC regime, as shown in Fig.~\ref{fig4}(b). Finally, negative $Z$ continues to support coherence peaks in the BCS regime while also significantly enhancing the sub-gap conductance in the BEC regime  [Fig.~\ref{fig4}(c)]. 
This enhancement is present within an energy interval about zero bias whose extent is limited by kinematic constraints imposed by the lead chemical potential $\mu_{\rm L}$ (the required backwards propagating hole state does not exist for $|E| > \mu_{\rm L}$). 

Our 1D analysis extends to the experimentally relevant scenario of a multi-channel lead normally incident on a 2D  
superconductor~\cite{SM}. In the limit of many ballistic channels, the total conductance $G(E)$ follows from an angular average over the contributions from all momenta in the plane of the interface~\cite{Kashiwaya1996, Daghero2010, Daghero2013, SM}. For isotropic $s$-wave superconductors this procedure leads to essentially identical results as in our 1D model. Nodal superconductors also show qualitatively similar ``revival" behavior at negative $Z$ in the BEC regime; the key distinguishing feature is that sub-gap conductance generated by Andreev reflection in the BCS regime acquires an inverted V-shape~\cite{Kashiwaya1996} due to transmission processes that suppress Andreev reflection away from zero bias~\cite{SM}. 

{\bf \emph{Discussion.}}~The suppression of Andreev reflection in a quantum point contact geometry, and its recovery by tunneling through potential wells,
clearly differentiates BEC phenomenology from the familiar properties of BCS superconductors.  In particular, the emergence of a pronounced $Z \rightarrow -Z$ asymmetry in the sub-gap conductance
provides a striking signature of BEC superconductivity.  
This feature stands in stark contrast to the above-gap conductance $G(E \gg \Delta)$ that is manifestly symmetric with respect to the sign of $Z$~\cite{SM}.  Possible experimental setups for probing the negative-$Z$ regime include (i)~STS measurements using a tip with an effective quantum well at its end, implemented by attaching, e.g., a quantum dot (reminiscent of single-electron transistor probes~\cite{Yoo1997,Martin2008}) or an impurity atom, or else by coating the tip with a layer of material with a smaller work function and (ii)~2D electronic transport setups where the N-S interface is tunable, either by applying a local gate near the junction or by constructing ``via contacts''  etched into different encapsulating insulators~\cite{Cao2022}.

Throughout the manuscript we focused on superconductivity arising from a single parabolic band. Extending these ideas to BEC physics in multiband superconductors, believed to be important to understanding recent experiments~\cite{Rinott2017}, is thus warranted. The limit of narrow-band superconductivity, where quantum geometry plays an important role in determining the superfluid stiffness~\cite{Peotta2015, Julku2016, Hofmann2020, Torma2021superfluidity, tian2021, Julku2021, Verma2021}, is another interesting avenue for future work with possible applications to moir\'e materials~\cite{Park2021, Kim2021}. Finally, self-consistent treatments~\cite{Mohit2022} and extensions beyond mean-field~\cite{Pistolesi1996} might lead to further insights into BEC superconductors and  associated pseudogap physics~\cite{deMelo1993, emery1995} at temperatures above $T_c$. 

\section{Acknowledgments}

We thank Mohit Randeria, Hyunjin Kim, Michał Papaj, and Kevin Nuckolls  for insightful discussions. This work was supported by the Gordon and Betty Moore
Foundation’s EPiQS Initiative, Grant GBMF8682 (C.L. and \'E.L.-H.); the Army Research Office under
Grant Award W911NF-17-1-0323; the Caltech Institute for Quantum
Information and Matter, an NSF Physics Frontiers Center with support of the Gordon and Betty Moore Foundation through Grant GBMF1250; and the Walter Burke
Institute for Theoretical Physics at Caltech.

\bibliography{references.bib}

\appendix
\setcounter{secnumdepth}{2} 
\renewcommand\thefigure{S\arabic{figure}}    
\setcounter{figure}{0}  
\section{BTK analysis of tunneling into a 2D superconductor}
\label{app:BTK_analysis}

Here we summarize the Blonder-Tinkham-Klapwijk (BTK) formalism~\cite{BTK1982} as applicable to tunneling into a 2D superconductor. The mathematical setup is analogous to that considered in the literature for the theory of anisotropic superconductors~\cite{Kashiwaya1996, Daghero2013}, where the barrier transparency parameter contains no geometrical factors. Formally, the problem reduces to a series of 1D BTK scattering problems as we explain below; correspondingly, along the way we will describe the derivation of 1D BTK results quoted in the main text. Unlike in the conventional BTK analysis, however, the Andreev approximation~\cite{andreev1964} is not taken to allow capturing both the BCS and BEC regimes.  

We will consider tunneling onto both gapped $s$-wave and nodal superconductors.  For a physical picture, imagine carving a hole into a 2D superconductor (with mass $m_{\rm SC}$, chemical potential $\mu_{\rm SC}$) and filling in the hole with a disc-shaped lead (with mass $m_{\rm L}$, chemical potential $\mu_{\rm L}$).  
An incident electron from the lead can then scatter into the adjacent 2D superconductor in any in-plane direction, with a given in-plane direction defining an effective 1D scattering problem as shown Fig.~\ref{fig1}c.  

Denoting the in-plane direction characterized by an angle $\theta$ as the $x$ axis for concreteness, we take our scattering states in the lead region ($x \le 0$) to be
\begin{align}
    \phi^L_e(x) &= e^{i q_e x} + a_e e^{-i q_e x} \, , \nonumber \\
    \phi^L_h(x) &= a_h e^{i q_h x}\,.
\label{eq:wave_states}
\end{align}
Here $q_e =  \sqrt{2m_{\rm L} (\mu_{\rm L}+E)} $ and $q_h =  \sqrt{2m_{\rm L} (\mu_{\rm L}-E)}$ are the momenta of the electron and hole, respectively, with $E$ the incident electron energy. We normalized the amplitude of the incoming electron plane wave to $1$, while $a_e$ and $a_h$ are the amplitudes associated with normal and Andreev reflection. 

In the superconducting region ($x > 0$) our ansatz takes the form
\begin{align}
    \phi^S_e(x) &= b_+ u_+ e^{i k_+ x} + b_- u_- e^{-i k_- x} \nonumber \\
    \phi^S_h(x) &= b_+ v_+ e^{i k_+ x} + b_- v_- e^{-i k_- x}\,,
    \label{SC_wavefunctions}
\end{align}
where $k_\pm = \sqrt{2 m_{\mathrm{SC}} \left(\mu_\mathrm{SC}\pm \sqrt{E^2-\Delta^2}\right)}$ are the momenta corresponding to the two solutions ($C$ and $D$) shown in Fig.~\ref{fig1}c. Coefficients $b_\pm$ correspond to the transmission of electrons into the superconductor without ($k_+$, process $D$) and with ($k_-$, process $C$) branch crossing. Possible angular dependence of the superconductor's order parameter enters through the pairing amplitude $\Delta(\theta)$, with $\theta$ the polar angle in the 2D plane. We note that in other geometries~\cite{Kashiwaya1996, Daghero2013} the angular dependence of the pairing amplitude $\Delta(\theta)$ need not be identical for the two momenta $k_+$, $k_-$. Following the BTK notation~\cite{BTK1982} we introduced our scattering states using the standard coherence factors $u_{\pm}$, $v_{\pm}$, evaluated at their respective branches, that here follow from: 
\begin{align}\label{eq:bcs_coherence_factors}
    u_+^2 &= \frac{1}{2} \left(1+\frac{\sqrt{E^2-\Delta^2}}{E}\right) = v_-^2 \,,\\
    v_+^2 &= \frac{1}{2} \left(1-\frac{\sqrt{E^2-\Delta^2}}{E}\right) = u_-^2\,.
\end{align}
While the above formulation is conventional and is used for a finite bias analysis, for analytical treatment of the zero-bias conductance it helps to define the scattering states without the $E=0$ nominal divergence; see next section. We stress that for the $s$-wave case angular dependence drops out, while for the nodal (e.g., $d$-wave) case angular dependence enters only via the pairing potential $\Delta(\theta)$.

To determine the scattering coefficients we include a $\lambda \delta(x)$ potential as in the main text and employ standard boundary conditions at $x=0$,
\begin{align}
\label{eq:bc_conditions}
    &\phi^L_e(x=0)=\phi^S_e(x=0)\,, \phi^L_h(x=0)=\phi^S_h(x=0)\,,\\
    &\frac{1}{2 m_\mathrm{SC}} \left.\frac{d \phi^S_e}{d x}\right\rvert_{x=0} - \frac{1}{2 m_\mathrm{L}} \left.\frac{d \phi^L_e}{d x}\right\rvert_{x=0} = \lambda \phi^S_e(x=0)\,,\\
    &\frac{1}{2 m_\mathrm{SC}} \left.\frac{d \phi^S_h}{d x}\right\rvert_{x=0} - \frac{1}{2 m_\mathrm{L}} \left.\frac{d \phi^L_h}{d x}\right\rvert_{x=0} = \lambda \phi^S_h(x=0)\,.
\end{align}
The first line ensures continuity of the wavefunction, while the other two lines impose a discontinuity in the wavefunction's first derivative as appropriate for the discontinuous change in mass and $\delta$-function potential at $x = 0$. The resulting system of equations is solved numerically for a given choice of parameters.

To maintain unitarity of the scattering problem and correctly obtain probabilities for Andreev reflection and normal reflection (i.e., probabilities $A$ and $B$ in the main text) one must amend the amplitudes $a_e, a_h, b_\pm$ with factors that account for the ratio of group velocities between the outgoing and incoming channels~\cite{BTK1982}.  This process ensures probability conservation for the four scattering processes such that $A+B+C+D= 1$, thus justifying use of the expression Eq.~\eqref{eq:buttiker} for the total junction conductance.  We thereby obtain 
\begin{align}
\label{eq:group_velocities}
&A = |a_h|^2 \left\lvert \frac{v_\mathrm{A}}{v_{\mathrm{IN}}}\right\rvert\,, &B = |a_e|^2 \left\lvert \frac{v_\mathrm{B}}{v_{\mathrm{IN}}}\,\right\rvert, \\
&C = |b_-|^2 \left\lvert \frac{v_\mathrm{C}}{v_{\mathrm{IN}}}\right\rvert\,, &D = |b_+|^2 \left\lvert \frac{v_\mathrm{D}}{v_{\mathrm{IN}}}\, \right\rvert,
\end{align}
with $v_\mathrm{A/B/C/D}$ the respective group velocities of carriers at momenta $q_h, q_e, k_\pm$. Note that the incoming group velocity is $v_{\mathrm{IN}} = v_\mathrm{B}$, and in the sub-gap regime $C,D=0$ is enforced by the vanishing group velocity for the evanescent solutions.  The conductance then follows as
\begin{align}
\label{eq:G_anisotropic}
    G(E) &= \int_0^{2 \pi} \frac{d\theta}{2\pi} G(E, \theta) \\
    &= \int_0^{2 \pi} \frac{d\theta}{2\pi} [1+A(E,\theta)-B(E,\theta)]\,,
\end{align}
where we made the angular dependence of the probability coefficients evident. For the case of an isotropic pairing where $\Delta(\theta) = \Delta$, the above angular integral is trivial, and the expression reduces to the form of Eq.~\eqref{eq:buttiker} considered in the main text.

In our modeling we further add a tiny imaginary part to the energy $E \rightarrow E + i \Gamma$ \cite{PhysRevLett.41.1509}. This factor accounts for a finite lifetime due to thermal broadening/impurity scattering, while also regularizing the solution at $E=\Delta$ [where the BTK scattering problem, cf.~Eqs.~\eqref{eq:bc_conditions}, becomes underdetermined].

Lastly, we comment on the behavior of the conductance for $E \ge \mu_{\rm L}$ in the BEC regime as seen in Fig.~\ref{fig4}. Andreev processes are kinematically forbidden since hole states are unavailable in the lead at those energies given the parabolic band structure considered in our model. We thus find vanishing of the Andreev reflection probability ($A \to 0$) for $E_{\mathrm{gap}} \ge E \ge \mu_{\rm L}$. Here $E_{\mathrm{gap}} = \sqrt{\Delta^2+\Theta(-\mu_\mathrm{SC}) \mu_{\mathrm{SC}}^2}$ is the quasi-particle gap shown in Fig.~\ref{fig4}.

\subsection{BTK analysis of sub-gap conductance for the $s$-wave case}

Although the treatment from the preceding section can be directly applied to the sub-gap conductance for a gapped $s$-wave superconductor, it is insightful to examine the analytic structure of the scattering problem up front in this regime. We start by considering the 1D version of BTK formalism~\cite{BTK1982} in the sub-gap limit $|E| < \Delta$ where we can safely neglect quasiparticle transmission across the N-S interface. The only remaining processes are normal reflection (process $B$ in Fig.~\ref{fig1}c) as an electron and Andreev reflection as a hole (process $A$ in Fig.~\ref{fig1}c). 

We take as an ansatz plane-wave solutions for electrons and holes in the lead region ($x\leq 0$) as in Eqs.~\eqref{eq:wave_states}. In the superconducting region $x>0$ we now specialize to the sub-gap regime and replace Eqs.~\eqref{SC_wavefunctions} with
evanescent solutions of the form
\begin{align}
    \phi^{S}_e(x) &= b_e e^{-\kappa x}, \\
    \phi^{S}_h(x) &= b_h e^{-\kappa x},
\end{align}
where normalizability requires $Re[\kappa] > 0$. Solving the eigenvalue equation $H_{\rm BdG} \Psi = E \Psi$ with
\begin{align}
     H_{\rm BdG}(x) = \begin{pmatrix}
     -\frac{1}{2m_{\rm SC}} \partial_x^2 - \mu_{\rm SC} & \Delta \\
     \Delta^* & \frac{1}{2m_{\rm SC}} \partial_x^2 + \mu_{\rm SC}
     \end{pmatrix}
 \end{align}
identifies two possible solutions for $\kappa$, denoted $\kappa_\pm$, which satisfy
\begin{equation}
    \frac{\kappa_\pm^2}{2m_{\rm SC}} = - \mu_{\rm SC} \pm i \sqrt{|\Delta|^2 - E^2}.
    \label{kappa_pm}
\end{equation}
Upon finding the corresponding eigenvectors as well, the general solution for the electron and hole wavefunctions in the superconductor region reads
\begin{align}
    \phi^{S}_e(x) =& b_e^+ e^{-\kappa_+ x} + b_e^- e^{-\kappa_- x}, \\
    \phi^{S}_h(x) =& \frac{E + i \sqrt{|\Delta|^2 - E^2}}{\Delta} b_e^+ e^{-\kappa_+ x} \nonumber \\
    +&  \frac{E - i \sqrt{|\Delta|^2 - E^2}}{\Delta} b_e^- e^{-\kappa_- x}.
\end{align}
Note that the structure of the prefactors of $b_e^\pm$ differs from those in Eqs.~\eqref{SC_wavefunctions} and \eqref{eq:bcs_coherence_factors} in a manner that streamlines analysis of the sub-gap conductance. In particular, in the present form the $E\to 0$ limit can be taken straightforwardly.

Repeating the analysis of Eq.~\eqref{eq:bc_conditions} and specializing to the zero-bias $E=0$ limit, we find an Andreev reflection amplitude 
\begin{equation}
    a_h = \frac{ v_{\rm L} (v_- - v_+) }{v_{\rm L}^2 + v_+ v_- + 2 v_{\rm L} Z(v_+ + v_-) + 4 v_{\rm L}^2 Z^2}\,.
    \label{eq:appendix_Ah}
\end{equation}
Here we defined velocities $v_{\rm L} = q_F/m_{\rm L}$, $v_\pm = \kappa_\pm/m_{\rm SC}$ and barrier strength parameter $Z =  m_{\rm L} \lambda/q_F = \lambda/v_{\rm L}$. We also used the fact that at zero energy $q_e = q_h = \sqrt{2 m_{\rm L} \mu_{\rm L}} \equiv q_F$. Equation~\eqref{kappa_pm} reduces to $\kappa_\pm^2/2m = - \mu_{\rm SC} \pm i \Delta$ at zero energy. 
Upon decomposing $\kappa_\pm = \kappa_{\rm R} \pm i \kappa_{\rm I}$ as in the main text, Eq.~\eqref{eq:appendix_Ah} takes the more convenient form 
\begin{equation}
    a_h = \frac{ -2 i v_{\rm L} v_{\rm I}}{ (v_{\rm R} + 2 v_{\rm L} Z)^2 + v_{\rm I}^2 + v_{\rm L}^2 },
     \label{eq:appendix_Ah_v2}
\end{equation}
which leads to the zero-bias Andreev reflection probability $A_0 = |a_h|^2$ quoted in Eq.~\eqref{eq:BTK_finiteZ}. Unlike the case of finite-bias tunneling, here the group velocity of the incident electrons and Andreev reflected holes is identical; hence the group velocity factors in Eqs.~\eqref{eq:group_velocities} drop out.

\subsection{Effective barrier parameter $\tilde{Z}$ and tunneling in the $E\gg \Delta$ limit}

In the standard BTK tunneling problem it is assumed that the masses of the lead and the superconductor are identical. More generally, as noted in the main text [Eq.~\eqref{eq:tilde_Z}], different masses (or equivalently different group velocities) can be reabsorbed in the definition of an effective barrier parameter $\tilde{Z}$ \cite{Lee2019}:
\begin{align}
\label{eq:tilde_Z_app}
    \tilde{Z}^2 &= \frac{(Z + \beta/2)^2}{\alpha} + \frac{(\alpha-1)^2}{4 \alpha},
\end{align}
such that the zero-energy Andreev reflection probability reads %
\begin{equation}
\label{eq:A0_tilde_Z_app}
    A_0 = \frac{1}{\left(1 + 2\tilde{Z}^2\right)^2}.
\end{equation}
Here, as in the main text, $\alpha = v_{\rm I}/v_{\rm L}$ and $\beta = v_{\rm R}/v_{\rm L}$. The zero-bias conductance across the BCS to BEC evolution, in the presence of both an interface potential and Fermi velocity mismatch, can thus always be written in the general form Eq.~\eqref{eq:A0_tilde_Z_app} with a renormalized transparency parameter. These expressions are also at the root of the data collapse observed in the Andreev reflection spectra of various high-density (BCS-like) superconductors with vastly different barrier and velocity mismatch parameters~\cite{Lee2019}.

Maximizing Andreev reflection translates to making $\tilde{Z}$ as small as possible, which in turn requires that each contribution to $\tilde{Z}$ needs to be \emph{individually} small because they add in quadrature. In the BCS limit where $\beta$ can be neglected, optimal Andreev reflection occurs for $Z=0$ and $\alpha=1$. In the BEC limit $Z=-\beta/2$ is optimal---but we still need $\alpha \sim 1$ which in BEC superconductors translates to $v_{\rm L} \sim v_{\rm I} \approx \Delta/(m_{\rm SC} v_F)$.
Velocity matching thus requires that the Fermi velocity of the lead be small. This condition can be achieved  via a small lead chemical potential and/or a large lead effective mass since $v_{\rm L} = q_F/m_{\rm L} = \sqrt{(2 \mu_{\rm L})/(m_{\rm L})}$ (for a parabolic band). 
Commercially available nickel-based STM tips with $m_L \sim 28 m_e$ could serve as promising leads in this regard. As shown in App.~\ref{sec:multiple_channels_app}, however, the condition for observing the optimal tunneling regime $v_{\rm L} \sim v_{\rm I}$ becomes somewhat relaxed in the presence of multiple scattering channels (see Fig.~\ref{fig_2_W_supplement}).

The above generalized barrier parameter $\tilde{Z}$ manifests in the high-bias $G(E\gg \Delta)$ limit as well. For these large energies, the problem reduces to tunneling from a normal metal to a normal metal across a potential barrier (or well), with different group velocities in the two metals. In such a case the tunneling conductance can be written in the standard form
\begin{equation}
\label{eq:G_EgtrDelta_app}
    G(E\gg \Delta) \approx \frac{1}{1+\tilde{Z}_N^2}.
\end{equation}
The effective tunneling barrier parameter for normal-normal scattering, $\tilde{Z}_N$, follows from
\begin{align}
\label{eq:tilde_ZN_app}
    \tilde{Z}_N^2 &= \frac{Z^2}{\alpha_N} + \frac{(\alpha_N-1)^2}{4 \alpha_N},
\end{align}
where $\alpha_N = v_{\rm I, N}/v_{\rm L, N}$ and $v_{\rm I, N}=k_+/m_\mathrm{SC}$, $v_{\rm L, N} = q_e/m_\mathrm{L}$ (see App.~\ref{app:BTK_analysis}). The expression \eqref{eq:G_EgtrDelta_app} is manifestly $Z \to -Z$ symmetric and maximal at $Z=0$, demonstrating that the $Z \to -Z$ asymmetry is solely a property of the sub-gap conductance in the BEC regime (at least within our theoretical framework).

\section{Generalization to nodal pairing}

\begin{figure*}
	\includegraphics[width=\textwidth]{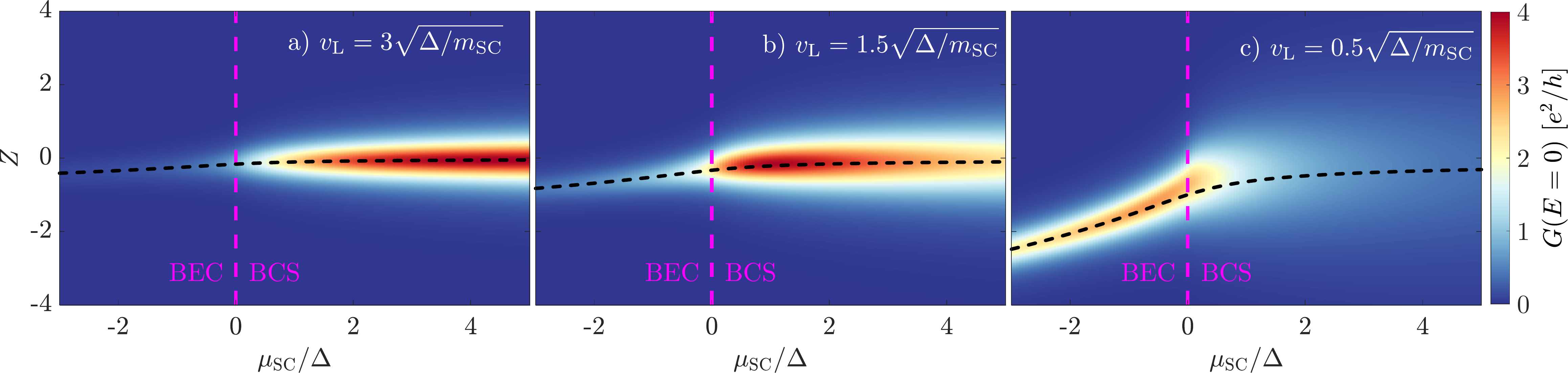}
	\caption{ Dependence of the BTK zero-bias tunneling conductance on the barrier parameter $Z$ along the BCS-BEC evolution for a $d$-wave superconductor ($\Delta(\theta) = \Delta \cos(2\theta)$), with $v_{\rm L} \sqrt{m_{\rm SC}/\Delta}$ set to (a) 3, (b) 1.5, and (c) 0.5.
	Black dashed lines trace $Z = -v_{\rm R}/2 v_{\rm L}$, one of the necessary conditions for perfect Andreev reflection. Incoming electrons tunnel through a potential barrier at $Z>0$ but a potential well at $Z<0$. While the conductance in the BCS regime is approximately $Z\rightarrow -Z$ symmetric, pronounced asymmetry emerges in the BEC regime, with enhanced Andreev reflection possible at negative $Z$ [see panel (c)]. Note the qualitative similarity with Fig.~\ref{fig2}, which displays analogous results for $s$-wave superconductors.}
	\label{fig2_d_wave}
\end{figure*}

\begin{figure*}
	\includegraphics[width=\textwidth]{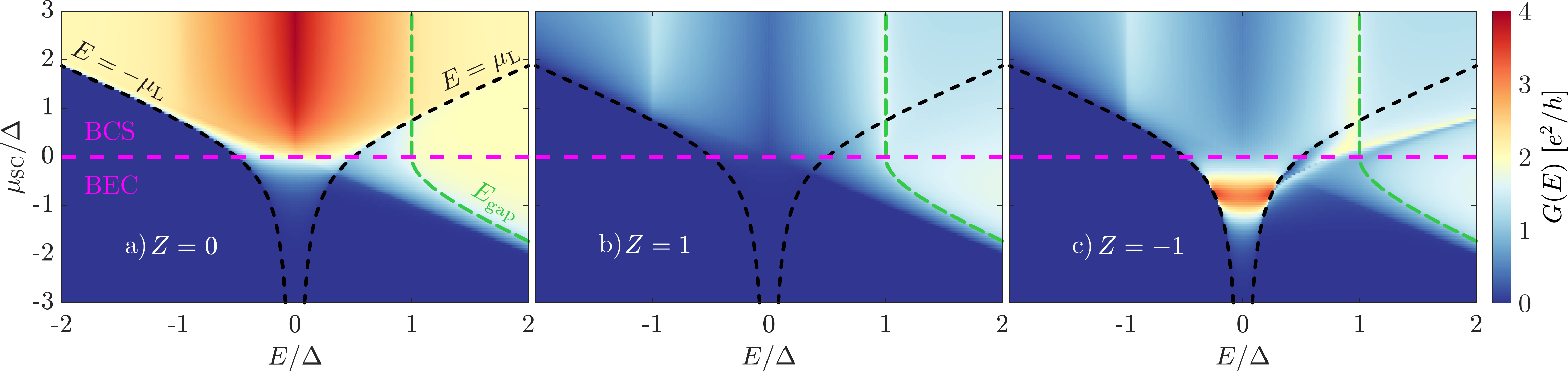}
	\caption{Finite-bias tunneling spectroscopy across the BCS-BEC evolution for a $d$-wave superconductor ($\Delta(\theta) = \Delta \cos(2\theta)$) with (a) $Z=0$, (b) $Z=1$ and (c) $Z=-1$, assuming an optimal lead 
	with $v_{\rm L} = v_{\rm I}$ for each $\mu_{\rm SC}$ value.
	Black dashed lines at energies $E = \pm \mu_{\mathrm L}$ show the corresponding optimal lead chemical potential $\mu_{\rm L}$, and green dashed lines denote the quasiparticle gap $E_\mathrm{gap}$ in the anti-nodal direction. (a) The quantum point contact limit $Z=0$ exhibits a sub-gap conductance enhancement due to Andreev reflection in the BCS regime, but not in the BEC regime. In contrast to $s$-wave superconductors [see Fig.~\ref{fig4}] the sub-gap conductance has an ``inverted V-shape" profile reflecting the presence of gapless excitations. (b,c) Tunneling spectra for $Z = \pm 1$ are similar in the BCS regime, showing a gap surrounded by coherence peaks at $E=\pm \Delta$. In the BEC regime, sub-gap conductance enhancement occurs \emph{only} for tunneling through potential wells, as in (c). Note the qualitative similarity with Fig.~\ref{fig4}, which displays analogous results for $s$-wave superconductors.}
	\label{fig4_d_wave}
\end{figure*}

For superconductors with anisotropic pairing gaps, we can compute the tunneling conductance---assuming a ballistic point contact such that the in-plane components of the incident electron momentum is conserved---by averaging over the in-plane momentum components as captured in Eq.~\eqref{eq:G_anisotropic}. (See also further discussion in Sec.~\ref{sec:multiple_channels_app}.)  Let us assume an isotropic normal state, such that the angular dependence of the Andreev and normal reflection coefficients only arise through the nodal pairing potential $\Delta(\bk) \sim \Delta_0 \cos( \nu \theta)$ with $\nu=1,2,3,...$ for $p$-wave, $d$-wave, $f$-wave, ... .

Figures~\ref{fig2_d_wave} and \ref{fig4_d_wave} show the $d$-wave analogs of the $s$-wave plots from the main text. The zero-bias conductance shown in Fig.~\ref{fig2_d_wave}a-c is qualitatively similar to that of Fig.~\ref{fig2}a-c, with the locus of perfect Andreev reflection tracing the $Z=- v_\mathrm{R}/2 v_\mathrm{L}$ condition as before. As expected, distinctions are present in the finite-bias tunneling profile shown in Fig.~\ref{fig4_d_wave}a-c. In particular, in the BCS regime for the QPC limit $Z=0$ (Fig.~\ref{fig4_d_wave}a), instead of a plateau profile present in the sub-gap conductance, here we see an inverted V-shape profile accounting for the presence of gapless excitations. Crucially, however, the qualitative features remain the same as in Fig.~\ref{fig4}: suppression of Andreev reflection for the $Z=0$ case (Fig.~\ref{fig4_d_wave}a) and a pronounced $Z \to -Z$ asymmetry with revival of Andreev reflection in the BEC regime for $Z < 0$ (Fig.~\ref{fig4_d_wave}b,c). At the level of this analysis we expect other pairing symmetries to produce similar tunneling profiles.

\section{ELF model and duality}

Here we explore the Effective Lead Model (ELF) model that complements the BTK analysis and provides an independent verification of the results. 
The lead is treated as a single-channel, semi-infinite wire defined along $x>0$, with $x=0$ proximate to an infinite $d$-dimensional superconductor (e.g., for $d=2$ the lead tunnels into the center of an infinite superconducting sheet oriented in the $y-z$ plane as illustrated in  Fig.~\ref{fig3}). It is convenient to equivalently view the lead as an infinite \emph{chiral} wire \cite{Fidkowski2012}: $x<0$ is then associated with \emph{incoming} states whereas $x>0$ corresponds to the \emph{outgoing} states.
With this viewpoint in mind, our starting point is the Hamiltonian
\begin{align}\label{eqn:ELF-ham-tot}
    H
    &=
    H_\mathrm{Lead}+H_\mathrm{SC}+H_{\mathrm{Lead}\text{-}\SC},
    \nt
    H_\mathrm{Lead}&=
    \int_x \psi^\dag(x)\left(-iv_\mathrm{L}\partial_x\right)\psi(x) + U_0\psi^\dag(0)\tau^z\psi(0)
    \nt
    H_\SC&=
    \int_{\br} d^\dag(\br)\Bigg[ 
    \left(-\frac{\boldsymbol{\partial}^2}{2m_\SC}-\mu_\SC\right)\tau^z
    +\Delta \tau^x\Bigg] 
    d(\br)
    \nt
    H_{\mathrm{Lead}\text{-}\mathrm{SC}}
    &=
    t[\psi^\dag(x=0)\tau^zd(\br = 0) +h.c.]\,.
\end{align}
Here, $\psi(x)$ and $d(\br)$ are Nambu spinors respectively associated with the lead and superconductor, i.e.,
\begin{align}
    \begin{pmatrix} d_1(\br)\\ d_2(\br)\end{pmatrix}
    &=
    \begin{pmatrix} c_\uparrow(\br)\\ c_\downarrow^\dag(\br)\end{pmatrix},
    &
    \begin{pmatrix} \psi_1(x)\\\psi_2(x)\end{pmatrix}
    &=
    \begin{pmatrix} \chi_\uparrow(x)\\ \chi_\downarrow^\dag(x)\end{pmatrix},
\end{align}
with $c_{\alpha}(\br)$ and $\chi_{\alpha}(x)$ the associated annihilation operators for electrons with spin $\alpha$.
Pauli matrices $\tau^{0,x,y,z}$ act on the Nambu indices.  The second line of Eq.~\eqref{eqn:ELF-ham-tot} incorporates the lead kinetic energy with velocity $v_{\rm L}$ as well as a potential $U_0$ at the interface with the superconductor.  The third describes the $d$-dimensional superconductor, which we assume has mass $m_{\rm SC}$, chemical potential $\mu_{\rm SC}$, and $s$-wave pairing amplitude $\Delta$. The final line incorporates spin-conserving tunneling between the lead at $x = 0$ and the superconductor at ${\bf r} = 0$ with strength $t$.  

Since the superconductor is fully gapped, we can integrate out its degrees of freedom to obtain the effective lead-only Hamiltonian
\begin{align}
    H_\mathrm{ELF}
    &=
    \int_z \psi^\dag(x)\left(-iv_\mathrm{L}\partial_x\right)\psi(x)
    \nt&\quad
    +
    \psi^\dag(0)\left[(U_0 + U_\mathrm{SC})\tau^z + 
    W_\mathrm{SC}\tau^x\right]\psi(0).
    \label{eq:H_ELF}
\end{align}
The new couplings introduced by the superconductor are given by
\begin{align}\label{eqn:app_ELF_param_def}
    U_\mathrm{SC}
    &=
   t^2\int_{\bk}\frac{\xi(\bk)}{\xi^2(\bk)+\Delta^2},
    \nt
    W_\mathrm{SC}
    &=
   t^2\int_{\bk} \frac{\Delta}{\xi^2(\bk)+\Delta^2},
\end{align}
where $\xi(\bk)=\bk^2/(2m_\SC) - \mu_\SC$ is the normal state dispersion of the superconductor.
Note that here we ignore all frequency dependence of these coefficients as well as any wavefunction renormalization, an
approximation that is justified by the low energies that we focus on.
Although we assumed $s$-wave pairing for convenience, a similar effective lead-only Hamiltonian follows also with nodal pairing symmetry in the BEC phase (where the superconductor is also gapped).  
The sole difference is that we would then include angular dependence of the pairing amplitude in Eq.~\eqref{eqn:app_ELF_param_def} via $\Delta\to\Delta(\bk)$.

Next we solve $H_{\rm ELF}$ to quantify the zero-bias conductance predicted by this formulation.  A wavefunction $\varphi(x)$ of energy $E$ must satisfy
\begin{align}
    E\varphi(x)
    &=
    -iv_\mathrm{L}\partial_x\varphi(x)
    +\delta(x)\left( U_\mathrm{eff}\tau^z+W_\mathrm{SC}\tau^x\right)\varphi(x),
\end{align}
where we have followed the main text and defined $U_{\rm eff} = U_0 + U_{\rm SC}$.  
Away from $x=0$, the solutions are plane waves with momentum $k=E/v_\mathrm{L}$.
The boundary conditions at $x=0$ are
\begin{align}
    \varphi(0^+)-\varphi(0^-)
    &=
    -\frac{i}{2v_\mathrm{L}}\left({U}_\mathrm{eff}\tau^z + {W}_\mathrm{SC}\tau^x\right)
    \nonumber \\
    &\times \left[\varphi(0^+)+\varphi(0^-)\right],   
\end{align}
where $0^\pm$ indicates $0\pm\varepsilon$ in the limit $\varepsilon\to0$, $\varepsilon>0$ and on the right-hand side we have regularized $\phi(0)$ symmetrically as $\left[\varphi(0^+)+\varphi(0^-)\right]/2$. The incoming and outgoing states for our semi-infinite lead of interest accordingly take the form
\begin{align}
    \varphi_\mathrm{in}(x)
    &=
    \begin{pmatrix}
    a_e\\ a_h
    \end{pmatrix}e^{ikx},
    \qquad
    x<0,
    \nt
    \varphi_\mathrm{out}(x)
    &=
    \begin{pmatrix}
    b_e\\b_h
    \end{pmatrix}e^{ikx},
    \qquad
    x>0\,,
\end{align}
with $a_{e/h}$ the incoming electron/hole amplitudes and $b_{e/h}$ the outgoing electron/hole amplitudes.
For our purposes it suffices to consider incoming electrons only and set $a_h=0$. 
In that case, the boundary conditions reduce to 
\begin{align}
    \begin{pmatrix}
    b_e-a_e\\b_h
    \end{pmatrix}
    &=
    -\frac{i}{2v_\mathrm{L}}
    \begin{pmatrix}
    {U}_\mathrm{eff} &   {W}_\mathrm{SC}\\
    {W}_\mathrm{SC} &   -{U}_\mathrm{eff}
    \end{pmatrix}
    \begin{pmatrix}
    b_e+a_e\\b_h
    \end{pmatrix}.
\end{align}
Solving the system of equations returns
\begin{align}
    \frac{b_e}{a_e}&=
    \frac{4v_\mathrm{L}(2v_\mathrm{L}-i{U}_\mathrm{eff})}{{U}_\mathrm{eff}^2+{W}_\mathrm{SC}^2+(2v_\mathrm{L})^2} - 1,
    \nt
    \frac{b_h}{a_e}&=
    -\frac{4iv_\mathrm{L} {W}_\mathrm{SC}}{ {U}_\mathrm{eff}^2+{W}_\mathrm{SC}^2+(2v_\mathrm{L})^2},
\end{align}
yielding an Andreev reflection coefficient 
\begin{align}
    A^\mathrm{ELF}&=\left|\frac{b_h}{a_e}\right|^2
    =
    \frac{4(2v_\mathrm{L})^2 {W}_\mathrm{SC}^2}{\left[{U}_\mathrm{eff}^2+{W}_\mathrm{SC}^2+(2v_\mathrm{L})^2\right]^2},
    \label{AELF}
\end{align}
which can be rewritten in the form of Eq.~\eqref{eq:ELF_andreev} in the main text.  Note that Eq.~\eqref{AELF} is independent of energy within the sub-gap regime where the effective lead-only Hamiltonian is applicable.  Energy independence reflects the fact the perturbations on the second line of Eq.~\eqref{eq:H_ELF} are marginal. Generically, additional perturbations at $x = 0$ involving derivatives of $\psi$ would be present and would yield nontrivial energy dependence not captured in our treatment.  Such corrections drop out, however, at zero energy---which we specialized to in the main text Eq.~\eqref{eq:ELF_andreev}.

Returning to the specific form of the perturbations in Eq.~\eqref{eqn:app_ELF_param_def}, for $d = 1$ the integrals evaluate to
\begin{align}
    {U}_\mathrm{SC} &= t^2\frac{\sqrt{m_\mathrm{SC}}}{2} \sqrt{\frac{-\mu_{\rm SC} + \sqrt{\Delta^2 + \mu_{\rm SC}^2}}{\Delta^2 + \mu_{\rm SC}^2}}
    \label{USC} \\
    {W}_\mathrm{SC} &= t^2\frac{\sqrt{m_\mathrm{SC}}}{2} \sqrt{\frac{\mu_{\rm SC} + \sqrt{\Delta^2 + \mu_{\rm SC}^2}}{\Delta^2 + \mu_{\rm SC}^2}}. 
    \label{WSC}
\end{align}
The duality between BCS and BEC superconductors is manifest here, as sending $\mu_{\rm SC} \rightarrow -\mu_{\rm SC}$ swaps $U_{\rm SC}\leftrightarrow W_{\rm SC}$.  Moreover, the explicit expressions above yield a ratio ${U}_\mathrm{SC} /{W}_\mathrm{SC} = v_\mathrm{R}/v_\mathrm{I}$---in agreement with the ELF/BTK correspondence provided in the main text below Eq.~\eqref{eq:ELF_andreev}.
(Interestingly, these relations persist despite the fact that the ELF model considers a slightly different geometry from BTK: the former tunnels into the center of an infinite 1D superconductor, while the latter tunnels into the end of a semi-infinite 1D superconductor.) 

\section{Tunneling with multiple channels}
\label{sec:multiple_channels_app}

Throughout the main text and the preceding supplemental material sections, we worked within the simplifying approximation of a single ballistic one-dimensional channel (or a sum of one-dimensional ballistic channels along the azimuthal directions in the case of 1D to 2D BTK geometry). Here we generalize our results to multiple scattering channels, showing that all conclusions of the main text remain valid. Moreover, we demonstrate that the condition for observing the optimal tunneling regime $v_{\rm L} \sim v_{\rm I}$---which requires a lead with a correspondingly small chemical potential and/or a large effective electron mass---becomes relaxed in the presence of multiple scattering channels.

\begin{figure*}
	\includegraphics[width=\textwidth]{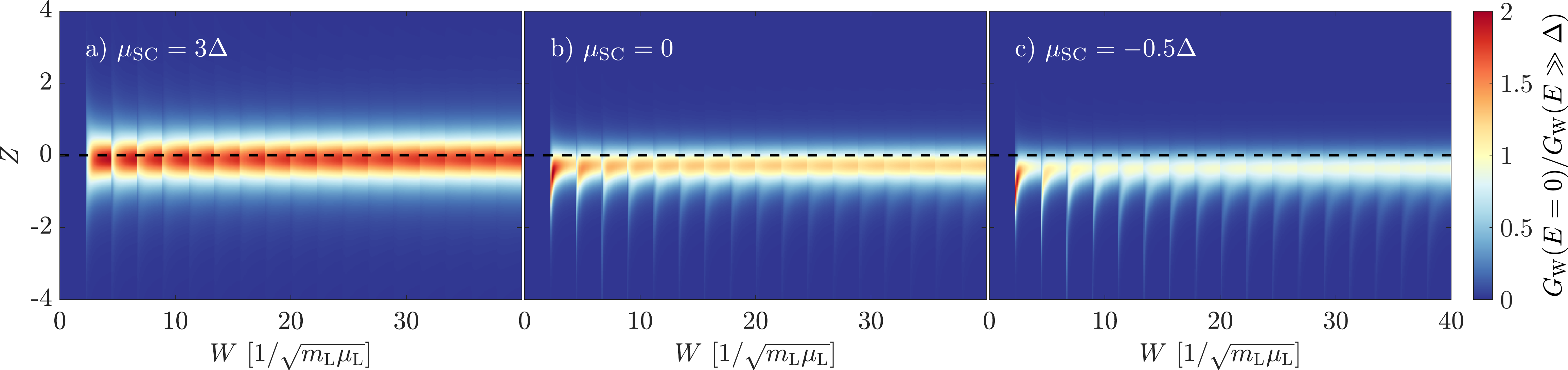}
	\caption{Normalized zero-bias conductance $G_\mathrm{W}(E=0)/G_\mathrm{W}(E\gg \Delta)$ of an $s$-wave superconductor as a function of lead width $W$ and tunneling parameter $Z$ for (a) BCS, (b) crossover, and (c) BEC regimes. Here the lead chemical potential is fixed for all three panels at $\mu_\mathrm{L}=3\Delta$ --- i.e., in contrast to Figs.~\ref{fig4} and \ref{fig4_d_wave} it is not tuned to its optimal value.}
	\label{fig_W_supplement}
\end{figure*}

\begin{figure*}
	\includegraphics[width=\textwidth]{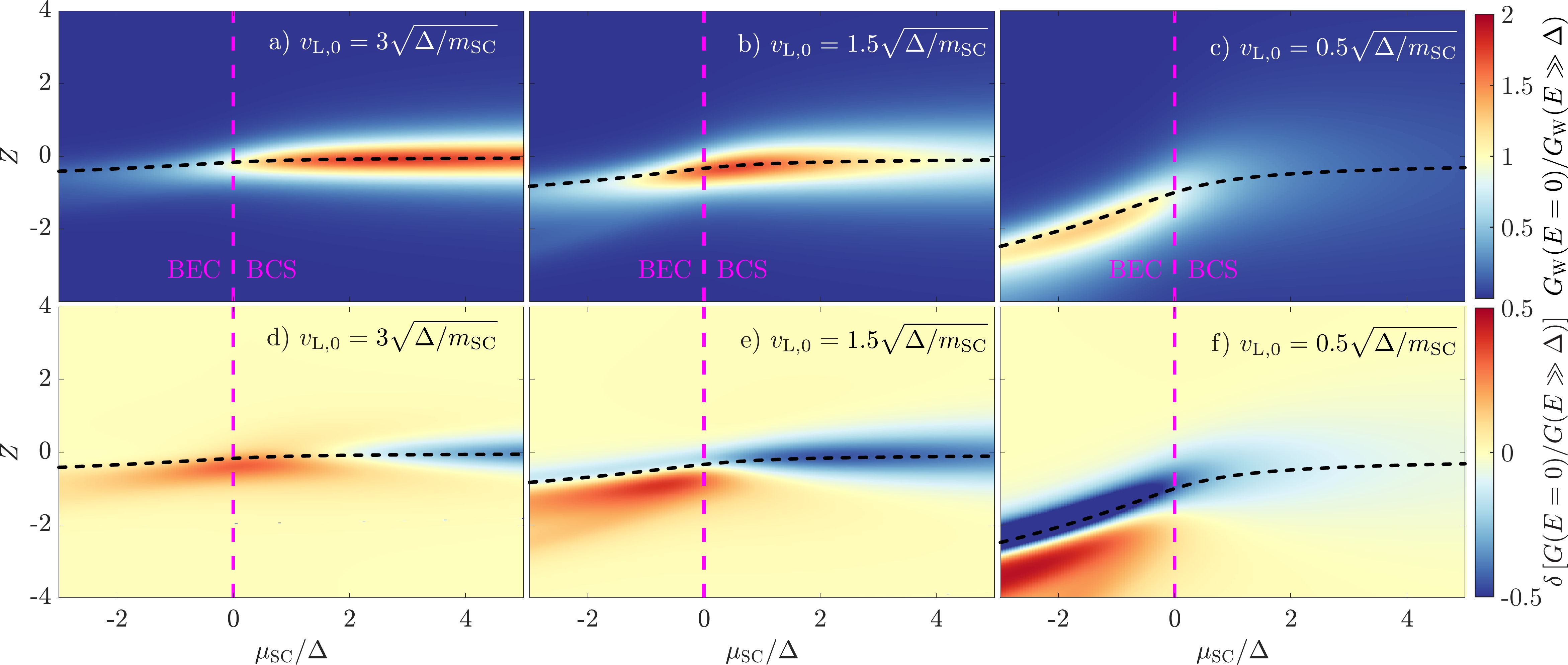}
	\caption{Dependence of the multi-channel BTK zero-bias tunneling conductance on the barrier parameter $Z$ along the BCS-to-BEC crossover, for a one-dimensional $s$-wave superconductor with $v_{{\rm L}, 0} \sqrt{m_{\rm SC}/\Delta}$ set to (a) 3, (b) 1.5, and (c) 0.5. A larger number of channels (Here we set $W = 40 \sqrt{2}/m_\mathrm{L} v_\mathrm{L,0}$) enlarges the parameter space where a non-zero signal is seen compared to Fig.~\ref{fig2}, but overall suppresses the maximum response as in Fig.~\ref{fig_W_supplement}. 
	Black dashed lines indicate the condition $Z = -v_{\rm R}/2 v_{\rm L}$ necessary for perfect Andreev reflection. In panels (d)-(f) we compute the difference between the panels of Fig.~\ref{fig2} (normalized by the appropriate above-gap conductance) and panels (a)-(c). In the BCS limit there is an overall reduction in the conductance, but in the BEC regime the range of $(Z,\mu_\mathrm{SC})$ parameters over which non-zero signal is seen is broadened.} 
	\label{fig_2_W_supplement}
\end{figure*}

To develop a model of multiple-channel scattering, we consider a quasi-1D lead with width $W$ along $y$ and semi-infinite extent along $x$. Assuming an isotropic 2D dispersion $\xi_{\bq} = q^2/2m_{\rm L}$ for the lead, the band structure of the quasi-1D problem exhibits sub-bands associated with the quantized transverse momenta $q_y = n \pi/W$:
\begin{equation}
    \xi_{n}(q_x) = \frac{q_x^2}{2 m_{\rm L}} + \frac{n^2 \pi^2}{2m_{\rm L} W^2}.
    \label{eq:xi_subbands}
\end{equation}
The sub-band index $n$ runs from 1 to $n_c$, where $n_c$ is the largest integer that admits a momentum on the Fermi surface; that is, $n_c = \lfloor q_F W / \pi \rfloor$. 

The total conductance is then taken as a sum over the contributions from all such sub-bands (or channels),
\begin{equation}
    G_W(E=0) = \frac{4e^2}{h} \sum_{n=1}^{n_c} A_n
    \label{eq:Asum_multichannel}
\end{equation}
with (at zero bias)
\begin{equation}
    A_n = |a^{(n)}_h|^2 =
    \frac{ 4 v_{{\rm L}, n}^2 v_{\rm I}^2 }{\left[\left(2 \lambda + v_{\rm R}\right)^2 + v_{\rm I}^2 + v_{{\rm L}, n}^2\right]^2}.
    \label{eq:An}
\end{equation}
Compared to Eq.~\eqref{eq:BTK_finiteZ} we have replaced $v_{\rm L} \rightarrow v_{{\rm L}, n}$ to describe the Fermi velocity along the $x$ direction (perpendicular to the interface) of each sub-band $n$ of the lead; the sub-band dispersion relation in Eq.~\eqref{eq:xi_subbands} yields
\begin{equation}
\label{eq:vl_discrete_vals}
    v_{{\rm L}, n} = \frac{q^x_{\rm F, n}}{m_{\rm L}} = \sqrt{ \frac{2}{m_{\rm L}} \left( \mu_{\rm L} - \frac{n^2 \pi^2}{W^2} \right)}.
\end{equation}
Additionally, we replaced the factor $ v_\mathrm{L} Z$ with $\lambda$ to explicitly highlight lack of dependence of the tunnel barrier on the sub-band index $n$.

In Fig.~\ref{fig_W_supplement}a-c we show the zero-bias conductance $G_W(E=0)$ of a lead with fixed chemical potential, in (a) the BCS regime, (b) at the crossover, and (c) in the BEC regime. As the total conductance grows linearly with the number of channels, we normalize $G_W(E=0)$ by the above-gap (i.e., normal-state) conductance
\begin{equation}
\label{eq:normal_conduction_w}
    G_W(E\gg \Delta) = \frac{2 e^2}{h} \sum_{n=1}^{n_c} \frac{1}{1+\tilde{Z}_{N,n} ^2}\,,
\end{equation}
where $\tilde{Z}_{N,n}$ for normal-normal scattering is defined by Eq.~\eqref{eq:tilde_ZN_app}, and now depends explicitly on the sub-band index $n$ through $v_{\rm L} \rightarrow v_{{\rm L}, n}$. 

We recover all qualitative conclusions of the single-channel problem examined in the main text, including suppression of Andreev reflection for $Z=0$ in the BEC regime and its revival for $Z < 0$. The large number of channels manifests in an enlarged parameter range ($Z$, $\mu_\mathrm{SC}$) over which revival of Andreev reflection is seen, see Fig.~\ref{fig_2_W_supplement} and the comparison with Fig.~\ref{fig2} for the same parameter range.

\subsection{Scattering in 3D setups}

Finally, as a last generalization of our analysis we briefly comment on the experimentally relevant case of a 3D material in a quasi-1D lead geometry---e.g., a semi-infinite wire (or STM tip) along the $x$ direction with \emph{two} quantized transverse directions along $y$ and $z$, with $q_y = n\pi/W$ and $q_z = m\pi/W$. We imagine tunneling into the surface of a 3D superconductor. 

Taking the ``continuum" limit where the number of sub-bands is large ($W \gg 1/q_F$), we can transform the sum over sub-band indices $m,n$ into an integral over the subset of the lead Fermi surface with positive $q_x$ components (a half-sphere of forward-propagating modes) using
\begin{align}
    \sum_{m,n=1}^{n_c} &= \left( \frac{W}{\pi} \right)^2 \int_{\rm FS} dq_y dq_z \\
    &= \frac{W^2 q_F^2}{2\pi^2} \int_0^{\pi/2} d\theta \sin \left( 2 \theta \right) \int_0^{2 \pi} d\phi ,
\end{align}
with $\theta$ the angle measured from the $q_x$ axis and $\phi$ the azimuthal angle. The non-standard Jacobian $\frac{1}{2} q_F^2 \sin(2 \theta)$ arises from our hybrid coordinate system $(q_x, \theta, \phi)$.

To compute the conductance we substitute $v_{{\rm L}, n} \rightarrow v_{\rm L} \cos \theta$ in the expression for the Andreev reflection coefficient. For an s-wave superconductor with an isotropic Fermi surface, $A(\theta)$ therefore does not depend on the azimuthal angle. Taking into account the conservation of momentum in the plane of the interface, the pairing gap $\Delta$ entering the Andreev reflection coefficient can however become $\phi$ dependent, as in the case of nodal superconductors. We therefore have
\begin{equation}
    G(E) = \frac{4e^2}{h}\frac{W^2 q_F^2}{2 \pi^2} \int_0^{\pi/2} d\theta \sin\left( 2 \theta \right) \int_0^{2 \pi} d\phi  A(\theta,\phi)\,.
\end{equation}
Even in the case of an isotropic Fermi surface in the superconductor, $A$ can depend on $\varphi$ and $\theta$ via the possible angular and momentum dependence of the pairing amplitude $\Delta$, respectively.

The $\phi$ integral is analogous to the angular average used to compute the Andreev reflection for two-dimensional nodal superconductors in Appendix~\ref{app:BTK_analysis}. The presence of an additional dimension however manifests through the $\theta$ integral, which will modulate the numerical results shown in Figs.~\ref{fig2_d_wave} and \ref{fig4_d_wave}. A more detailed analysis of the conductance in such setups will thus depend on the details of the Fermi surfaces of both the lead and superconductor, as well as on the precise geometry of junction~\cite{Daghero2013}.

\end{document}